\documentclass[aps,prl,reprint,longbibliography,superscriptaddress]{revtex4-1}

 \usepackage{amsmath,bm}
 \usepackage{mathrsfs}
 \usepackage{amsfonts}
 \usepackage{graphicx}
 \usepackage{setspace} 
 \usepackage{graphicx}
 \usepackage{epstopdf}
 \usepackage{dcolumn}
 \usepackage{amsmath}
 \usepackage{epsfig}
 \usepackage{indentfirst}
 \usepackage{psfrag}
 \usepackage{subfigure}
 \usepackage{amssymb}
 \usepackage{color}
 \usepackage{units} 

 \usepackage{graphicx}
 \usepackage{dcolumn}
 \usepackage{bm}
 \usepackage{natbib}

\usepackage{physics}
\usepackage{dcolumn}
\usepackage{bm}
\usepackage{natbib}
\usepackage[backref=none,bookmarksnumbered=true,bookmarks=true,bookmarksopen=true,colorlinks=true,
citecolor=blue,linkcolor=blue,anchorcolor=green,urlcolor=blue,unicode=false]{hyperref}
\usepackage{atbegshi}

\definecolor{mypink1}{rgb}{0.858, 0.188, 0.478}
\def\NH2{NH$_2$}
\newcommand{\Figref}[1]{Fig.~\ref{#1}}

\newcommand{\rev}[1]{{\color{black} #1}}
\newcommand{\revi}[1]{{\color{black} #1}}
\newcommand{\agl}[1]{\textcolor{black}{#1}}
\newcommand{\np}[1]{\textcolor{black}{#1}}

\begin{document}
\title{\rev{Band Depopulation of Graphene Nanoribbons \\ Induced by Chemical Gating with Amino Groups}}

\author{Jingcheng Li}
\affiliation{CIC nanoGUNE \revi{BRTA}, Tolosa Hiribidea 76, 20018 Donostia-San Sebastian, Spain}

\author{Pedro Brandimarte}
\affiliation{Donostia International Physics Center (DIPC), 20018 Donostia-San Sebasti\'an, Spain}

\author{Manuel Vilas-Varela}
\affiliation{Centro Singular de Investigaci\'on en Qu\'imica Biol\'oxica e Materiais Moleculares (CiQUS), and Departamento de Qu\'imica Org\'anica, Universidade de Santiago de Compostela,\rev{15782 Santiago de Compostela}, Spain}

\author{Nestor Merino-\rev{D\'iez}}
\affiliation{CIC nanoGUNE \revi{BRTA}, Tolosa Hiribidea 76, 20018 Donostia-San Sebastian, Spain}
\affiliation{Donostia International Physics Center (DIPC), 20018 Donostia-San Sebasti\'an, Spain}

\author{C\'esar Moreno}
\affiliation{Catalan Institute of Nanoscience and Nanotechnology (ICN2), CSIC and \rev{The Barcelona Institute of Science and Technology, Campus UAB, Bellaterra, 08193 Barcelona} Spain}

\author{Aitor Mugarza}
\affiliation{Catalan Institute of Nanoscience and Nanotechnology (ICN2), CSIC and \rev{The Barcelona Institute of Science and Technology, Campus UAB, Bellaterra, 08193 Barcelona} Spain}
\affiliation{ICREA Instituci\'o Catalana de Recerca i Estudis Avancats, \rev{08193} Barcelona, Spain}

\author{Jaime S\'aez Mollejo}
\affiliation{Donostia International Physics Center (DIPC), 20018 Donostia-San Sebasti\'an, Spain}

\author{Daniel S\'anchez Portal}
\affiliation{Centro de F{\'{\i}}sica de Materiales 	MPC (CSIC-UPV/EHU),  20018 Donostia-San Sebasti\'an, Spain}
\affiliation{Donostia International Physics Center (DIPC), 20018 Donostia-San Sebasti\'an, Spain}

\author{Dimas G. de Oteyza}
\affiliation{Centro de F{\'{\i}}sica de Materiales 	MPC (CSIC-UPV/EHU),  20018 Donostia-San Sebasti\'an, Spain}
\affiliation{Donostia International Physics Center (DIPC), 20018 Donostia-San Sebasti\'an, Spain}
\affiliation{Ikerbasque, Basque Foundation for Science, 48013 Bilbao, Spain}

\author{Martina Corso}
\affiliation{Centro de F{\'{\i}}sica de Materiales 	MPC (CSIC-UPV/EHU),  20018 Donostia-San Sebasti\'an, Spain}
\affiliation{Donostia International Physics Center (DIPC), 20018 Donostia-San Sebasti\'an, Spain}

\author{Aran Garcia-Lekue}
\affiliation{Donostia International Physics Center (DIPC), 20018 Donostia-San Sebasti\'an, Spain}
\affiliation{Ikerbasque, Basque Foundation for Science, 48013 Bilbao, Spain}
\email{wmbgalea@ehu.eus}

\author{Diego Pe{\~{n}}a}
\affiliation{Centro Singular de Investigaci\'on en Qu\'imica Biol\'oxica e Materiais Moleculares (CiQUS), and Departamento de Qu\'imica Org\'anica, Universidade de Santiago de Compostela,\rev{15782 Santiago de Compostela}, Spain}
\email{diego.pena@usc.es}

\author{Jose Ignacio Pascual }
\affiliation{CIC nanoGUNE \revi{BRTA}, Tolosa Hiribidea 76, 20018 Donostia-San Sebastian, Spain}
\affiliation{Ikerbasque, Basque Foundation for Science, 48013 Bilbao, Spain}

\email{ji.pascual@nanogune.eu}


\begin{abstract} 
The electronic properties of graphene nanoribbons (GNRs) can be precisely tuned by chemical doping. Here we demonstrate \rev{that amino (NH$_2$) functional groups attached at the edges of chiral GNRs (chGNRs) can efficiently gate the chGNRs and lead to the valence band (VB) depopulation on a metallic surface.} The NH$_2$-doped chGNRs are grown by on-surface synthesis on Au(111) using functionalized \revi{bianthracene} precursors. Scanning tunneling spectroscopy resolves that the NH$_2$ groups significantly up-shift the bands of chGNRs, causing the Fermi level crossing of the VB onset of chGNRs. Through density functional theory simulations we confirm that the hole-doping behavior is due to an upward shift of the bands induced by the edge \revi{NH$_2$} groups. 

\end{abstract}
 
 \maketitle

\np{Graphene nanoribbons (GNRs) have emerged as a promising material  for nanoelectronics \cite{Llinas2017} since they combine many of the extraordinary properties of the parent graphene \cite{Han2014a,Moreno199}  with a high tunability of their electronic band structure\cite{Yazyev2013}. Contrary to graphene, GNRs are frequently semiconducting  materials, thus offering excellent perspectives for their utilisation as electronic components such as diodes \cite{Kargar2009} or transistors \cite{Schwierz2010,Bennett2013,Llinas2017,Moreno199}. 
A fascinating aspect of narrow GNRs is that their electronic properties can be  tuned
through the precise control of their atomic structure, which can be achieved thanks to the recent development of bottom-up on-surface synthesis (OSS) strategies \cite{Talirz2016,Clair2019}. 
These strategies rely on the careful design of suitable molecular precursors with specific shape and chemical composition to steer a step-wise reaction on a metal surface, leading to extended and atomically precise GNRs. 
An ample library of precursors and reaction pathways has been constructed in the last years, incorporating successful examples of precise control over the GNR's width\cite{cai2010,Chen2013,ABDURAKHMANOVA2014,Basagni2015,Kimouche2015,Zhang2015a,Talirz2017,Merino-Diez2017}, orientation and edge topology\cite{Han2014b,Liu2015,Ruffieux2016,DeOteyza2016,Rizzo2018,Groning2018,Merino-Diez2018, Moreno19}, which resulted in large variations of their electronic structure.     
Furthermore, OSS strategies allow to easily combine several precursors on a surface for  fabricating  complex structures such as atomically sharp graphene heterojunctions \cite{Cai2014,Chen2015},  quantum dots\cite{Carbonell-Sanroma2017,Wang2017}, or hybrid systems composed of metal-organic molecules embedded in GNRs \cite{Lieaaq0582,su2018,Li2019}, demonstrating their potential  role for GNRs based nanoelectronics.}

\rev{Since most GNRs are semiconducting, a promising method for tuning their band structure  is the electrostatic gating effect induced by doping \cite{Krull2013}.    Chemical doping of GNRs has been achieved through  modification of the molecular precursors to either incorporate  substitutional heteroatoms in the carbon backbone   \cite{Cai2014,Bronner2013,Kawai2015,Cloke2015,Nguyen2016,Carbonell-Sanroma2017,Durr2018,carbonell-sanroma_JPCC2018,Wang2018,Kawai2018} or by adding functional groups at the edges \cite{Carbonell-Sanroma2017a}. However, the band structure of the \revi{substitutionally}-doped systems \revi{could} not always be simply related to the pristine one (CB)\cite{Kawai2015,Cloke2015,Carbonell-Sanroma2017,Durr2018,carbonell-sanroma_JPCC2018}; the embedded heteroatoms have a profound impact on the bands' symmetries and character \cite{Cao2017}.
Instead, chemical functionalization of the GNR edges promisingly behaved as an efficient method to fine-tune the electronic structure  of pristine GNRs.
In the most simple scenarios, the effect of attached chemical groups can be described as an effective electrostatic gating of the native band structure \cite{Cai2014,Carbonell-Sanroma2017a,Wang2018}. The magnitude of edge chemical gating is small, to date barely restricted to  sub-gap down-shifts of the band structure.   
}

\begin{figure*}[hbt]
\includegraphics[width=1.7\columnwidth]{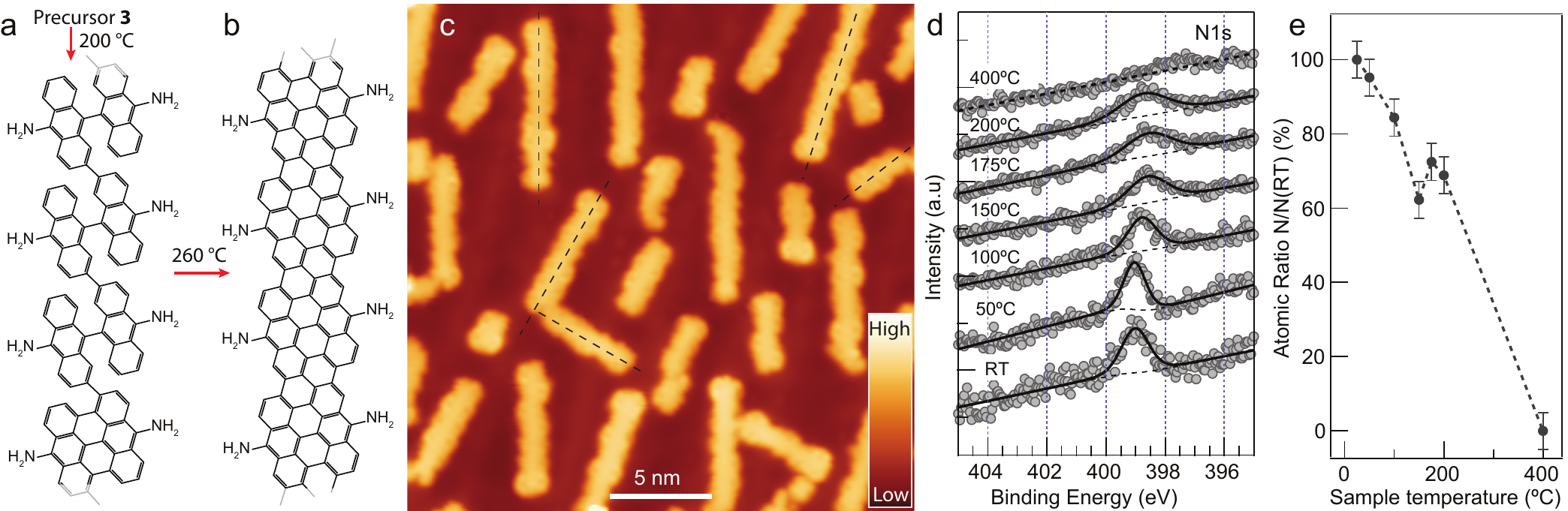}
\caption{Synthesis of NH$_2$-chGNRs. (a),(b) Reaction scheme of the polymerization and cyclodehydrogenation of molecular precursors respectively. For simplicity, only one chirality of NH$_2$-chGNRs is shown here.  (c) STM overview image of NH$_2$-GNRs grown on Au(111), taken at V$_s$  = 0.2 V, I$_t$  = 20 pA. The dashed lines show the preferred orientations of NH$_2$-chGNRs on the substrate. \rev{Same color scale is used for all the STM images.} (d,e) Characterization of  \revi{NH$_2$} groups at different annealing temperatures by N1s peaks in XPS measurements.  }
\label{fig:synthesis}
\end{figure*}
\rev{Here, we report on the very efficient chemical gating  of amino\revi{(NH$_2$)} functional groups attached at the edges of  narrow \revi{chiral GNRs with a sequence of 3 zigzag and 1 armchair sites ((3,1) chGNRs)}. The electron donating character of NH$_2$ is  inherited  by the GNR, resulting in an effective GNR charging by hole injection and valence band depopulation on a Au(111) surface.}
Experimentally, NH$_2$ end groups were substituted at the edges of pristine chGNRs through bottom-up synthesis with modified precursors. Our  results satisfactorily demonstrate that the electropositive affinity of the dopant is transmitted to the chGNR carbon backbone  \cite{Carbonell-Sanroma2017a}, which on a Au(111) surface becomes strongly hole doped. Through a combination of scanning tunnelling microscopy (STM) and spectroscopy (STS), X-ray photoelectron spectroscopy (XPS) measurements and density functional theory (DFT) calculations we show that  the  \revi{NH$_2$} groups induce a substantial  up-shift of the GNR bands, and cause their VB to cross the Fermi level.  The effective depopulation of the VB depends on the number of NH$_2$ groups per unit cell surviving the reaction, as well as on their relative alignment. 

\subsection{Results/Discussion}

\begin{figure}[b]
\includegraphics[width=0.95\columnwidth]{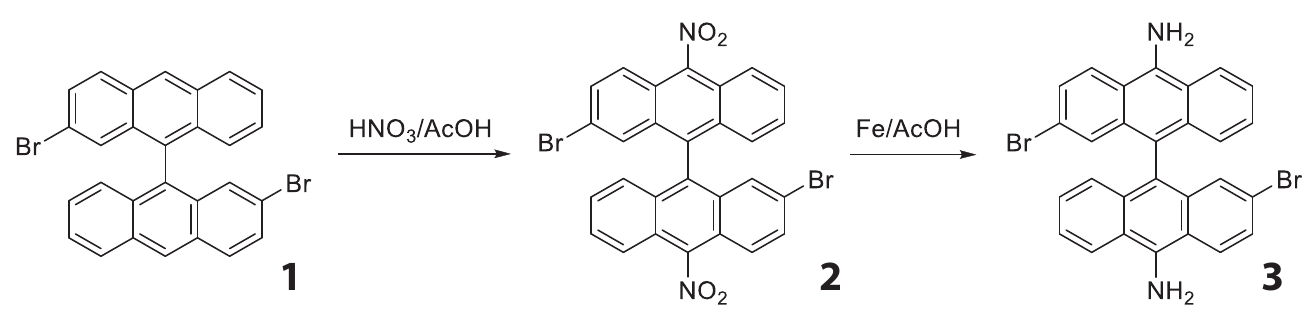}
\caption{Synthetic route to obtain 2,2'-dibromo-[9,9'-bianthracene]-10,10’-diamine (\textbf{3}).}
\label{fig:synthesis-route}
\end{figure}

\textbf{Synthesis of the molecular precursor:} To  functionalize (3,1) chGNRs with   \revi{NH$_2$} groups at the edge (named as NH$_2$-chGNRs), we prepared  the molecular precursor 2,2'-dibromo-[9,9'-bianthracene]-10,10’-diamine (\textbf{3}, \Figref{fig:synthesis-route}) 
in two steps from 2,2'-dibromo-9,9'-bianthracene (\textbf{1}). First, selective double nitration of compound \textbf{1} led to the synthesis of compound \textbf{2}, which was isolated with 64\% yield. Then, Fe-promoted reduction of bianthracene \textbf{2} in acetic acid afforded the GNR precursor \textbf{3} with 98\% yield (see Supporting Information for details).

The molecular precursor \textbf{3} was sublimated from a Knudsen cell  at 212 $^{\circ}$C  onto a clean Au(111) substrate kept at room temperature. The precovered sample was gradually annealed first to 200 $^{\circ}$C to induce the polymerization of compound \textbf{3} through Ullmann-like coupling  reactions (\Figref{fig:synthesis}(a)). Subsequently, the sample temperature was further raised to 260 $^{\circ}$C to  trigger the cyclodehydrogenation step causing the planarization of the polymers into NH$_2$-chGNRs (\Figref{fig:synthesis}(b)). \Figref{fig:synthesis}(c) shows a STM overview image of the resulting NH$_2$-chGNRs on a Au(111) surface. \rev{The doped chGNRs, as the pristine ones, display preferential orientations on the substrate,} but notably shorter than pristine chGNRs\cite{DeOteyza2016,Merino-Diez2018}, reaching lengths of up to 15 nm. \rev{We speculate that the attachment of  \revi{NH$_2$} groups lowers the diffusion of precursors and thus  hinders their polymerization processes\cite{Bronner2017}}. The edges of the resulting NH$_2$-chGNRs are quite inhomogeneous, and their width is larger than pristine chGNRs, suggesting the presence of a certain amount of  \revi{NH$_2$} groups attached to the edges (see \revi{Supporting} Information). 

To prove the survival of  \revi{NH$_2$} groups during the on-surface synthesis (OSS) process, we characterized the thermal evolution during the OSS with X-ray photoelectron spectroscopy (XPS) measurements (\Figref{fig:synthesis}(d,e)). We find a gradual decrease of the N-$1s$ XPS peak intensity with increasing annealing temperatures, which completely vanishes at 400$^{\circ}$C. Since the cyclodehydrogenation step completing the synthesis of chGNRs occurs at temperatures slightly above 200$^{\circ}$C \cite{DeOteyza2016,Merino-Diez2018}, the XPS results suggest that about 60\% of the \revi{NH$_2$} groups survive the OSS.

\begin{figure*}[t!]
\includegraphics[width=1.75\columnwidth]{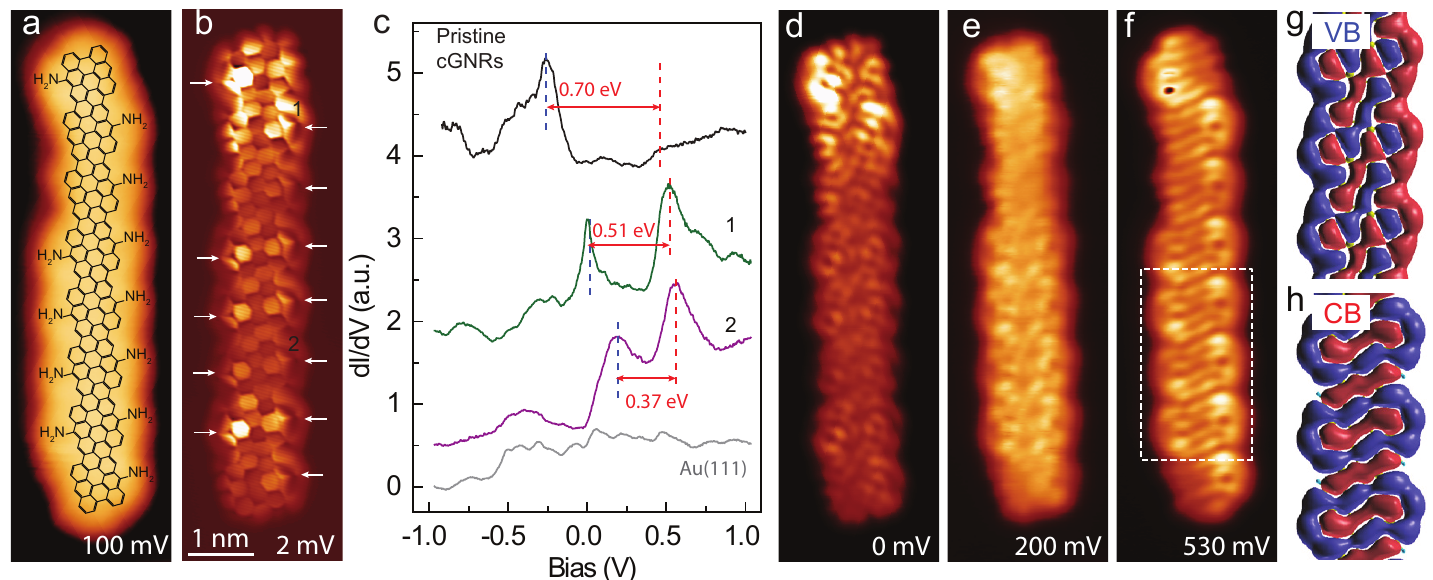}
\caption{Characterization of the atomic and electronic structures of NH$_2$-chGNRs. (a) Constant current image of a single NH$_2$-chGNRs with its \revi{chemical} structure superimposed on top. (b) Constant height current image (Vs = 2 mV) of the same NH$_2$-chGNRs in \revi{(a)} taken with CO-terminated tip. (c) dI/dV spectra taken on the locations as indicated with numbers in (\revi{b}). For comparison, dI/dV spectra taken on pristine chGNRs (black curve on top) are also shown here. Blue and red dashed lines indicate the energy locations of VB and CB, respectively. The values of the band gaps are noted in the figure. (Open-feedback parameters: Vs = 1 V, It = 1 nA; modulation voltage Vrms = 10 mV). (d-f) Constant height dI/dV maps of the ribbon in (a) taken at different bias values as indicated in the figures respectively (modulation voltage Vrms = 10 mV). All STM images share the same scale bar indicated in (b). (g),(h) DFT simulations of the wave function corresponding to states at the onset of VB and CB (at the $\Gamma$ point) of NH$_2$-chGNRs with all  \revi{NH$_2$} groups attached. Red and blue colors represent isosurfaces of positive and negative wave function amplitudes, for an isovalue of 0.015 \AA$^{-3/2}$.}
\label{fig:stmNH2chGNR}
\end{figure*}

\rev{\Figref{fig:stmNH2chGNR}(a) shows the constant current STM image of a NH$_2$-chGNR which shows irregular edges. The superimposed chemical structure of the NH$_2$-chGNR indicates the loss of some  \revi{NH$_2$} groups on the edges. To further prove this, we acquired high resolution STM image of the same GNR with a CO-terminated tip\cite{Gross2009,Kichin2011}, as shown in \Figref{fig:stmNH2chGNR}(b).} The high resolution image resolves the backbone structure of the ribbon as composed of zig-zag edged segments \cite{Merino-Diez2018}, and also shows distinctive triangular features at many of the zigzag edges (indicated by arrows in \Figref{fig:stmNH2chGNR}(b)), which we attribute to surviving  \revi{NH$_2$} groups as shown in \Figref{fig:stmNH2chGNR}(a). \rev{The carbon atoms with missing  \revi{NH$_2$} groups are then passivated by hydrogen atoms during the on-surface reactions\cite{Talirz2013}}. Most of the chGNR segments maintain at least one of the \NH2 edge groups attached at one side, and some appear with both of them intact.  
From such high resolution images \rev{on more than 80 GNRs}, we extract that about 75\% of the \NH2 groups survived the on-surface reaction. This value is even larger than the one obtained from the XPS results, \rev{most probably due to the different annealing processes (see methods).} \revi{This is a remarkable doping value when compared with the chemical doping surviving the reaction for cyano\revi{(CN)} functionalization of 7-AGNRs\cite{Carbonell-Sanroma2017a}. We note that edge functionalization  in such chiral ribbons is so efficient due to the lower temperatures required for their complete synthesis, in contrast with \textit{e.g.} 7-AGNRs, which need more than 300$^{\circ}$C for their OSS.}

\textbf{Electronic structure of doped chGNRs:}  \revi{NH$_2$} groups have an electron donating character. To study their effect on the electronic properties of the chiral nanoribbons, we recorded differential conductance plots (dI/dV) directly over singly and doubly NH$_2$-doped segments, and compare them with measurements on pristine chGNRs. As shown in \Figref{fig:stmNH2chGNR}(c), the characteristic 0.7 eV energy gap of pristine chGNRs \cite{Merino-Diez2018} is severely modified on the  \revi{NH$_2$}-substituted segments, which exhibit instead two pronounced peaks, indicated by red and blue dashed lines in \Figref{fig:stmNH2chGNR}(c). Over the singly doped segment, one of the peaks is pinned at the Fermi level (zero bias), while the other appears at \revi{500 mV} above. On the doubly-doped units, both peaks appear slightly shifted upwards, and their separation noticeably decreases. 

To elucidate the origin of the peaks, we acquired dI/dV maps at bias values corresponding to the peak energies, which unveil the spatial distribution of these states  (\Figref{fig:stmNH2chGNR}(d-f)). dI/dV maps at 0 mV and 200 mV (the bias values of the lower peaks in each case)  resemble the VB shape of pristine chGNRs at 
the singly- and doubly-doped segments, respectively. 
In contrast, the dI/dV map at 530 mV (in coincidence with the higher-bias peak in both cases) reproduces a wave-front pattern that is characteristic of the CB of pristine chGNRs\cite{Merino-Diez2018}. Furthermore, these dI/dV maps show very good agreement with the density functional theory simulations of wave function amplitude of valence and conduction bands along a fully NH$_2$-doped ribbon  (\Figref{fig:stmNH2chGNR}(g,h)).  Based on these observations, we ascribe the two dI/dV peaks  to the VB and CB onsets of NH$_2$-chGNRs, respectively. 

\begin{figure*}[t!]
\includegraphics[width=1.65\columnwidth]{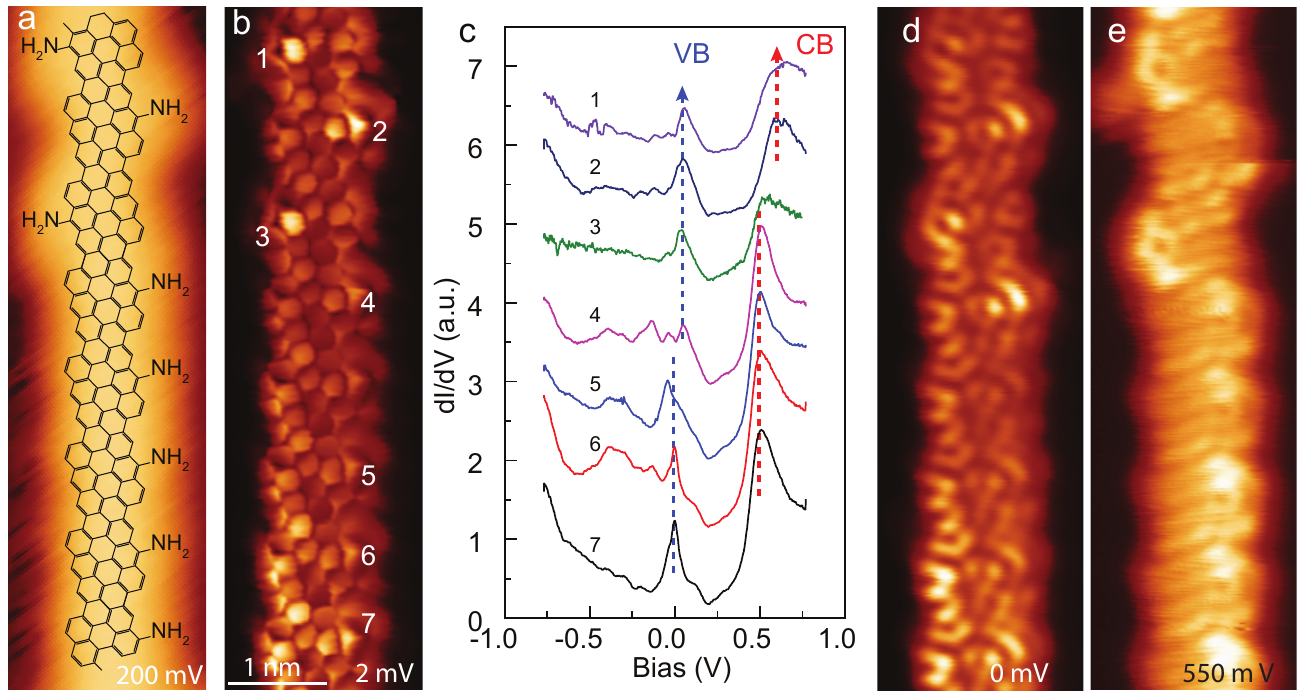}
\caption{Electronic structures of NH$_2$-chGNR with half  \revi{NH$_2$} groups lost.  (a) Constant current image of a single NH$_2$-chGNRs with its \revi{chemical} structure superimposed on top. (b) Constant height current image (Vs = 2 mV) of the same NH$_2$-chGNRs in \revi{(a)} taken with CO-terminated tip. (c) dI/dV spectra taken on the locations  indicated with numbers in (\revi{b}). Blue and red dashed lines indicate the energy onsets of VB and CB, respectively.  (Open-feedback parameters: Vs = \revi{1 V}, It = 1 nA, modulation voltage Vrms = 10 mV). (d-e) Constant height dI/dV maps of the ribbon in (a) taken at different bias values as indicated in the figures respectively (modulation voltage Vrms = 10 mV). All STM images share the same scale bar indicated in (b).}
\label{fig:halfNH2chGNR}
\end{figure*}  

Compared to pristine chGNRs, both VB and CB onsets of the  NH$_2$-doped segments appear shifted upwards, and the VB-CB energy gap is significantly reduced (as labelled in \Figref{fig:stmNH2chGNR}(c)). This upwards shift of the frontier bands can be attributed to a significant hole-doping introduced by the edge  \revi{NH$_2$} groups, whose effect increases with the number of  groups surviving in each segment. 
Particularly striking is the alignment of the VB onset with the Fermi-level for the singly-doped segments, and its further shift to $\sim$0.18 eV when a second \NH2 is present, unveiling a substantial band depopulation by edge-doping, and underpinning of chGNRs on Au(111).    

Intriguingly, the number of \NH2-dopants per unit is not the only parameter tuning the bands' onset, but also their relative arrangement. This was concluded from the inspection of ribbons with only one edge group per unit.  In order to induce the more diluted functionalization observed in Fig. 4, we annealed the sample
to a slightly higher temperature during the OSS process (\textit{e.g.} to 280 $^{\circ}$C for the data shown in Fig. 4). The result was a reduction of the edge functionalization to about 50\%, with a nice preference to maintain one group per unit cell rather than two or none.  We found that \NH2 groups could either line-up along the same edge (\textit{e.g.} as between segments 4 and 7 in Fig. 4(a,b), or alternate at opposite sides (as between 1 and 4 in Fig. 4(a,b)), what had an apparent impact in their corresponding band alignment. 
While the dI/dV spectra (\Figref{fig:halfNH2chGNR}(c)) and maps (\Figref{fig:halfNH2chGNR}(d) and (e)) appear in every case similar to the ones shown in \Figref{fig:stmNH2chGNR}, both VB and CB peaks in regions with alternating \NH2 groups (spectra 1 to 4 in \Figref{fig:halfNH2chGNR}(c)) appear  shifted upwards by $\sim$ 50 meV with respect to the regions where  \revi{NH$_2$} groups line up in the same side  (spectra 5 to 7).  
 
\begin{figure*}[t]
   \includegraphics[width=1.65\columnwidth]{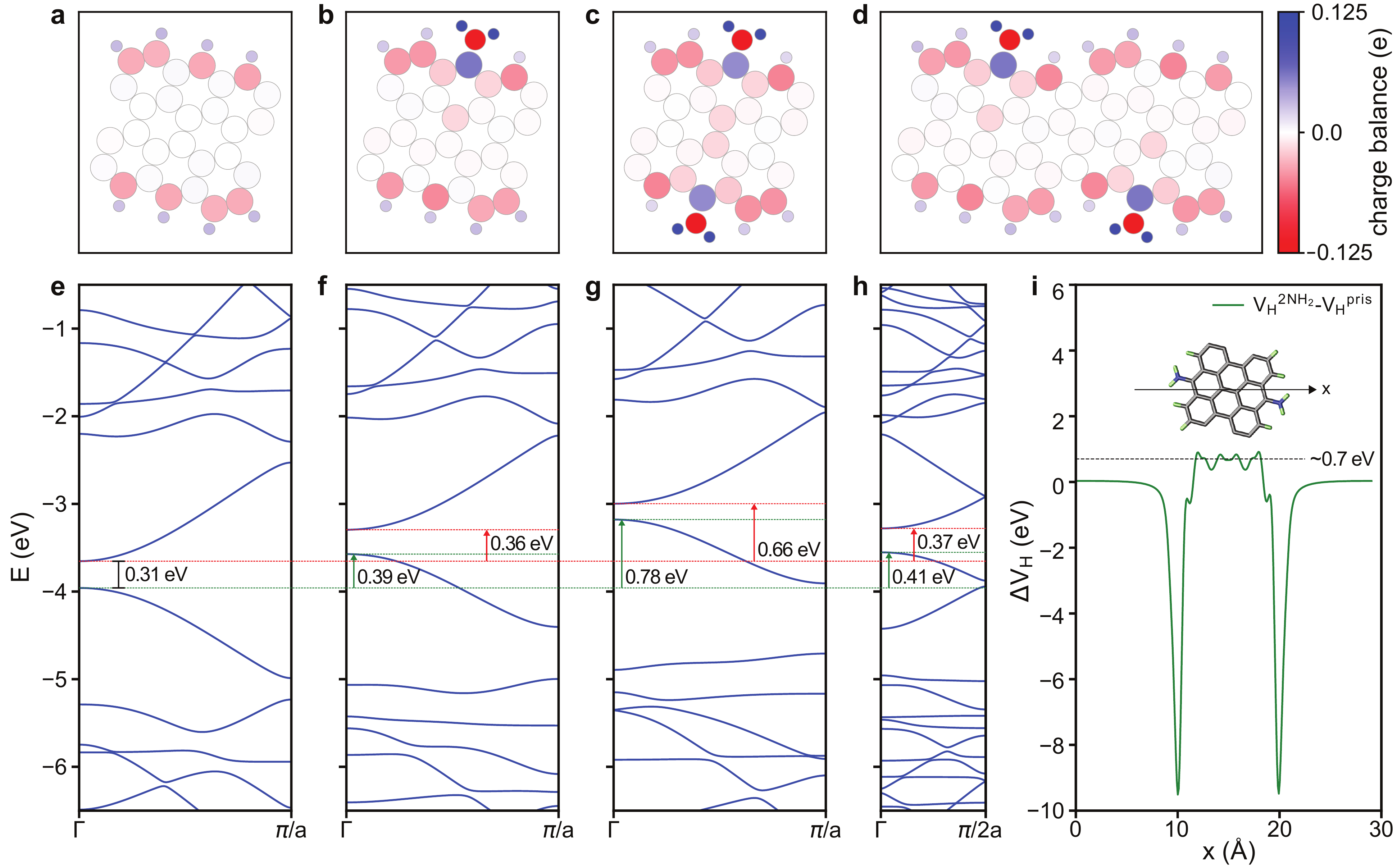}
   \caption{
   Simulated electronic structure and charges.
   Charge distribution calculated with a Hirshfeld population analysis for: (a) pristine chGNR, (b) chGNR with NH$_2$ groups on one side (1\revi{a}-NH$_2$-chGNR), (c) NH$_2$ groups on both sides (2-NH$_2$-chGNR), and (d) on alternating sides (\revi{1b}-NH$_2$-chGNR).
   Corresponding band structures obtained for (e) pristine chGNR, (f) 1a-NH$_2$-chGNR, (g) 2-NH$_2$-chGNR and (h) 1b-NH$_2$-chGNR.
   The shift in energy of the top of the VB and the bottom of the CB as compared to the pristine chGNR are marked in green and in red, respectively.
   (i) Electrostatic potential difference between the 2-NH$_2$-chGNR and a pristine chGNR, taken at the plane of the ribbon and averaged along the periodic direction (y).
   }
   \label{fig:dft}
\end{figure*}

\textbf{Theoretical analysis:} 
Further insight into the doping effects of  \revi{NH$_2$} groups is obtained by DFT calculations of free-standing nanoribbons (see Methods for details on the simulations). In particular, we studied the band structure of infinite (3,1) chGNRs with the edge structures shown in  \Figref{fig:dft}(a-d), namely, the three experimentally observed scenarios: NH$_2$ groups only on one side (1a-NH$_2$-chGNR), on both sides (2-NH$_2$-chGNR), and  in alternating positions (1b-NH$_2$-chGNR), as well as the pristine chGNRs for comparison. 
The corresponding band structures  (\Figref{fig:dft}(e-h)) show that, in all three cases, the presence of NH$_2$ groups at the edge induces an upward shift of the bands with respect to the pristine case.
\agl{For the singly-doped cases, the
VB band shifts about 0.4 eV. 
When both edges are doped (2-NH$_2$-chGNR), the 
upwards shift of the VB is almost doubled (0.78 eV), showing that it scales with the number of dopants.} 

The band structure also reproduces the clear reduction of the VB-CB energy gap with doping seen in the experiment. In particular, we find that the band gap of the singly-doped GNR decrease \np{$\approx$ 10\%} with respect to the pristine case, while this reduction amounts \np{to $\approx$ 40\%} for the doubly-doped species. 
A similar band gap closure with doping in GNRs was attributed to the  extension of the $\pi$-conjugated electron cloud outside the graphene backbone with the addition of the functional groups \cite{Carbonell-Sanroma2017a}. 
On the other hand, we note that the calculated shifts of the frontier bands are  larger than the measured ones, and the gap reduction obtained numerically is smaller than in experiments.
These  discrepancies might be related  to screening effects of the metallic substrate, which are not included in our calculations. Yet, our DFT results are in very good qualitative agreement with the experimental measurements. 

To explain the origin of the band shifts, we studied the  charge redistribution induced by the  \revi{NH$_2$} edge groups in the chGNRs. As seen from the Hirshfeld plots in  \Figref{fig:dft}(a-d), the  \revi{NH$_2$} groups redistribute their charges by the covalent bonding to the edge and induce a small electron depletion at the connecting C and accumulation right at the middle of the ribbon.
Interestingly, the lower electron density at the edge coincides with the higher current of these rings in the chGNR backbone (\Figref{fig:stmNH2chGNR}(a) and \Figref{fig:halfNH2chGNR}(a))).
We speculate that the higher current signal over the edge phenyl rings bonding to \NH2 and lower current over the neighbour rings at the center are related to the calculated electron depletion and accumulation, respectively, and \rev{to their influence on }the effective tunneling. 

As a result of the charge redistribution, a strong edge dipole of $\sim$3.6\,D per unit cell pointing  outwards from the ribbon backbone emerges around the NH$_2$ group.
Considering  the edge dipole per unit cell of the pristine chGNR ($\sim$2.1\,D), and since both CH and NH$_2$ dipoles are  pointing in the same direction, we can estimate that each NH$_2$ group contributes with a dipole of about 2.0\,D.
 This edge dipole  increases the internal electrostatic potential, and consequently, shifts  their occupied bands upwards. For example, for the 2-NH$_2$-chGNR case (\Figref{fig:dft}(i)), we find that the $\sim$0.7 eV higher electrostatic potential  is the main responsible of the $\sim$0.7 eV  up-shift of its VB. The result is that the NH$_2$ groups endow their electron-donating  character to the ribbons. On the Au(111) surface, this leads to a substantial hole-doping from the metal, reaching the point where the VB is partially depopulated. We note that this scenario differs from the previously reported case of armchair GNRs doped with CN groups. CN induced a dipole pointing opposite, \textit{i.e.}\ towards the ribbon's backbone, leading to a total dipole smaller in magnitude and opposite in sign to the one presented here. This caused a small band down shift, contrary to the situation  found here,  but the semiconducting character remained unchanged \cite{Carbonell-Sanroma2017a}.

\section{Conclusions}
\rev{Our results  demonstrate a chemical-gating strategy of graphene nanoribbons, by which  incorporating electron donating chemical groups to the edge of GNRs causes a significant hole-doping  on a Au(111) surface. As found previously with  \revi{CN} groups \cite{Carbonell-Sanroma2017a}, the GNR inherits the electron affinity of the attached chemical group. Favored by the lower reaction temperatures for the synthesis of chiral GNRs, we found that a large fraction of  \revi{NH$_2$} groups survived the on-surface reaction,  and caused the depopulation of its VB, signaling its charging. }   Based on DFT simulations, we explained the enhanced electron-donating character of the doped ribbons as arising from a substantial charge redistribution induced by the dopant. From these results, we foresee that combination of edge terminations with different electron affinity  character can be a promising route to engineer hybrid GNRs with active electronic properties.  

We acknowledge financial support from: i)  AEI/FEDER-EU through grants no. MAT2016-78293-C6, FIS2017-83780-P (AEI/FEDER,EU), the  Maria de Maeztu unit of excellence MDM-2016-0618, and the Severo Ochoa program (ICN2) SEV-2017-0706; ii) the European Research Council (grant agreement no. 635919); iii) the Xunta de Galicia (Centro singular de investigaci\'on de Galicia, accreditation 2016−2019, ED431G/09); iv) the EU project SPRING (863098); v) the European Regional Development Fund (ERDF) under the program Interreg V-A Espa\~na-Francia-Andorra (Contract No. EFA 194/16 TNI),  vi)  the CERCA Program/Generalitat de Catalunya, and vii) the Gobierno Vasco-UPV/EHU (project IT1246-19).  

\section{Methods/Experimental}

The experiments were performed on a low temperature (5 K) STM under ultrahigh vacuum (UHV) conditions. The Au(111) substrate was cleaned in UHV by repeated cycles of Ne$^{+}$ ion sputtering and
subsequent annealing to 460 $^{\circ}$C. The molecular precursor was sublimated at 210 $^{\circ}$C from a Knudsen cell onto the clean Au(111) substrate kept at room temperature. Then the sample was first annealed at 200 $^{\circ}$C for 15 minutes and then annealed at 260 $^{\circ}$C for 5 minutes. A tungsten tip  functionalized with a CO molecule was used for high resolution images. The high resolution images were acquired in constant height mode, at very small voltages, and junction resistances of typically 20 M$\Omega$. The $dI/dV$ signal was recorded using a lock-in amplifier with a bias modulation of $V_\mathrm{rms}=\revi{10}$ mV at 760 Hz.

XPS measurements were carried out using a Specs PHOIBOS 150 hemispherical energy analyzer using a monochromatic X-ray source (Al K$\alpha$ line with an energy of 1486.6 eV and 400 W) and energy referenced to the Fermi level. Surface cleanliness was confirmed with x-ray photoelectron spectroscopy (XPS) prior molecules deposition. Organic molecules were sublimated in UHV from a home-made Knudsen-cell type evaporator with an alumina crucible, with the sample kept at room temperature. \rev{Then the sample was step-wisely annealed to 400  $^{\circ}$C. The sample was kept at each temperature step in Fig. 2d for 10 minutes for the XPS measurements.}

\textbf{Computational Procedures}

The optimized geometry and electronic structure of free-standing pristine and NH$_2$ doped ch-GNRS were calculated using density functional theory, as implemented in the SIESTA code.\cite{Soler_2002}
The nanoribbons were relaxed until forces on all atoms were $<$ 0.01 eV/\AA, and the dispersion interactions were taken into account by the nonlocal optB88-vdW functional.\cite{Klime__2009}
The basis set consisted of double-$\zeta$ plus polarization (DZP) orbitals for all species, with an energy shift parameter of 0.01 Ry.
A 1$\times$1$\times$100 Monkhorst-Pack mesh was used for the k-point sampling of the Brillouin zone, where the 100 k-points are taken along the direction of the ribbon.
A cutoff of 300 Ry was used for the real-space grid integrations.
The atomic population analysis was performed using the Hirshfeld scheme for partitioning the electron density.\cite{hirshfeld_1977}
The dipole moments were calculated integrating the valence electron density at each point times the distance of that point to the ribbon backbone (central axis). The corresponding neutralizing contribution due to the nuclei is also added. More details can be find in  \citealt{Carbonell-Sanroma2017a}.

 \bibliographystyle{apsrev4-1} 
%

\end{document}


\title{Supplementary Information}

\newpage 

\section{Molecular precursor synthesis and characterization}

All reactions were carried out under argon using oven-dried glassware. TLC was performed on Merck silica gel 60 F254; chromatograms were visualized with UV light
(254 and 360 nm). Flash column chromatography was performed on Merck silica gel 60
(ASTM 230- 400 mesh). $^1$H and $^{13}$C NMR spectra were recorded at 300 and 75 MHz
(Varian Mercury 300 instrument) or 500 and 125 MHz (Bruker AMX-500 instrument),
respectively. Low-resolution electron impact mass spectra were determined at 70 eV on
a HP-5988A instrument. High-resolution mass spectra (HRMS) were obtained on a
Micromass Autospec spectrometer. 2,2'-Dibromo-9,9'-bianthracene (\textbf{1}, Figure S1) was
prepared following published procedures.\cite{DeOteyza2016} Commercial reagents and anhydrous
solvents were purchased from ABCR GmbH and Aldrich Chemical Co. and were used
without further purification.

\subsection{Synthesis of 2,2’-dibromo-10,10’-dinitro-9,9’-bianthracene}

\begin{figure}[htb!]
\includegraphics{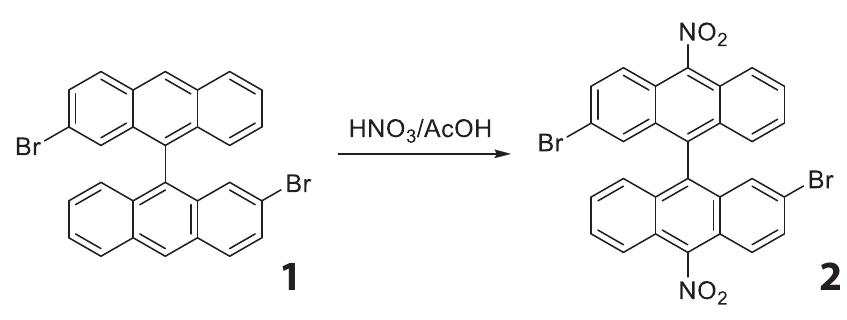}
\caption{Synthesis of 2,2’-dibromo-10,10’-dinitro-9,9’-bianthracene (2) }                                                                                 
\end{figure}

To a solution of compound \textbf{1}\cite{DeOteyza2016} (50 mg, 0.10 mmol) in AcOH (5 mL), a solution of HNO$_3$ (2
mL, 62\% in H2O) was added and the resulting mixture was stirred at room temperature
for 16 h. Then, a solution of HCl was added (5 mL, 10\% in H$_2$O) and the precipitate was
filtered. The solid was purified by column chromatography (SiO$_2$; hexane/CH$_2$Cl$_2$ 2:1) to
obtain compound \textbf{2} (38 mg, 64\%) as an orange solid (p.f. 384 $^\circ$C). $^1$H NMR (300 MHz,
CDCl$_3$) $\delta$: 8.07 (d, J = 9.1 Hz, 2H), 7.97 (d, J = 9.4 Hz, 2H), 7.76 - 7.67 (m, 4H), 7.39 - 7.27 (m, 2H), 7.23 (d, J = 1.2 Hz, 2H), 7.01 (d, J = 8.7 Hz, 2H) ppm. $^{13}$C NMR (75 MHz,
CDCl3) $\delta$: 145.8 (2C), 133.7 (2C), 132.7 (2CH), 131.5 (2C), 131.3 (2C), 129.4 (2CH),
128.2 (2CH), 127.8 (2CH), 126.3 (2CH), 123.7 (2CH), 122.5 (2C), 122.3 (2C), 122.0
(2CH), 120.5 (2C) ppm. MS (EI) m/z (\%): 602 (M+, 100), 572 (17), 348 (24), 337 (23),
174 (27), 169 (26). HRMS (EI): C$_{28}$H$_{14}$N$_2$O$_4$Br$_2$; calculated: 599.9320, found: 599.9322.

\subsection{Synthesis of 2,2'-dibromo-[9,9'-bianthracene]-10,10’-diamine}
\begin{figure}[h]
\includegraphics{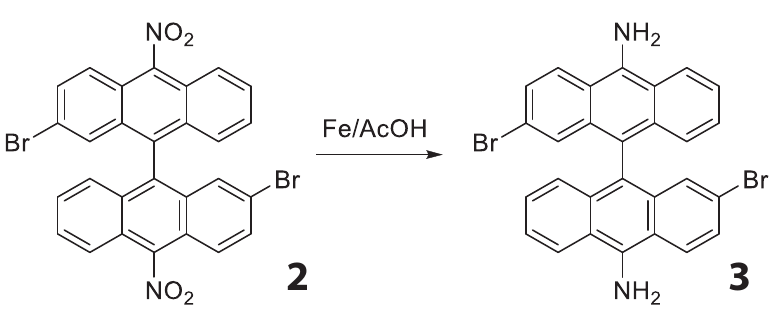}
\caption{Synthesis of 2,2'-dibromo-[9,9'-bianthracene]-10,10’-diamine (3)}                                                                                 
\end{figure}

A mixture of compound \textbf{2} (25 mg, 0.042 mmol) and Fe (70 mg, 1.25 mmol) in AcOH (5
mL) was stirred at room temperature for 16 h. Then, H$_2$O (10 mL) and AcOEt (10 mL)
were added, the organic layer was separated and washed with aqueous solution of
NaOH (10\%, 5 mL). The organic layer was dried over Na$_2$SO$_4$, filtered and evaporated
under reduce pressure. The crude product was purified by column chromatography
(SiO$_2$; CH$_2$Cl$_2$) to obtain compound \textbf{3} (22 mg, 98\%) as an orange solid (p.f. 337 ºC). $^{1}$H
NMR (500 MHz, CD$_3$OD) $\delta$: 8.34 (d, J = 8.9 Hz, 2H), 8.27 (d, J = 9.4 Hz, 2H), 7.36 - 7.33
(m, 4H), 7.11 (ddd, J = 8.8, 6.4, 1.2 Hz, 2H), 7.06 (d, J = 2.0 Hz, 2H), 6.95 (d, J = 8.7 Hz,
2H), 4.87 (s, 4H) ppm. $^{13}$C NMR (125 MHz, CD3OD) $\delta$: 142.6 (2C), 134.7 (2C), 134.6
(2C), 129.4 (2CH), 128.0 (2CH), 127.3 (2CH), 126.8 (2CH), 126.0 (2CH), 124.4 (2CH),
123.6 (2CH), 121.3 (2C), 120.6 (2C), 119.5 (2C), 117.2 (2C) ppm. MS (EI) m/z (\%): 542
(M+, 100), 271 (14), 190 (10), 176 (14). HRMS (EI): C$_{28}$H$_{18}$N$_2$Br$_2$; calculated: 539.9837,
found: 539.9833.

\subsection{Characterization by $^{1}$H and $^{13}$C NMR}

\begin{figure}[h]
\includegraphics{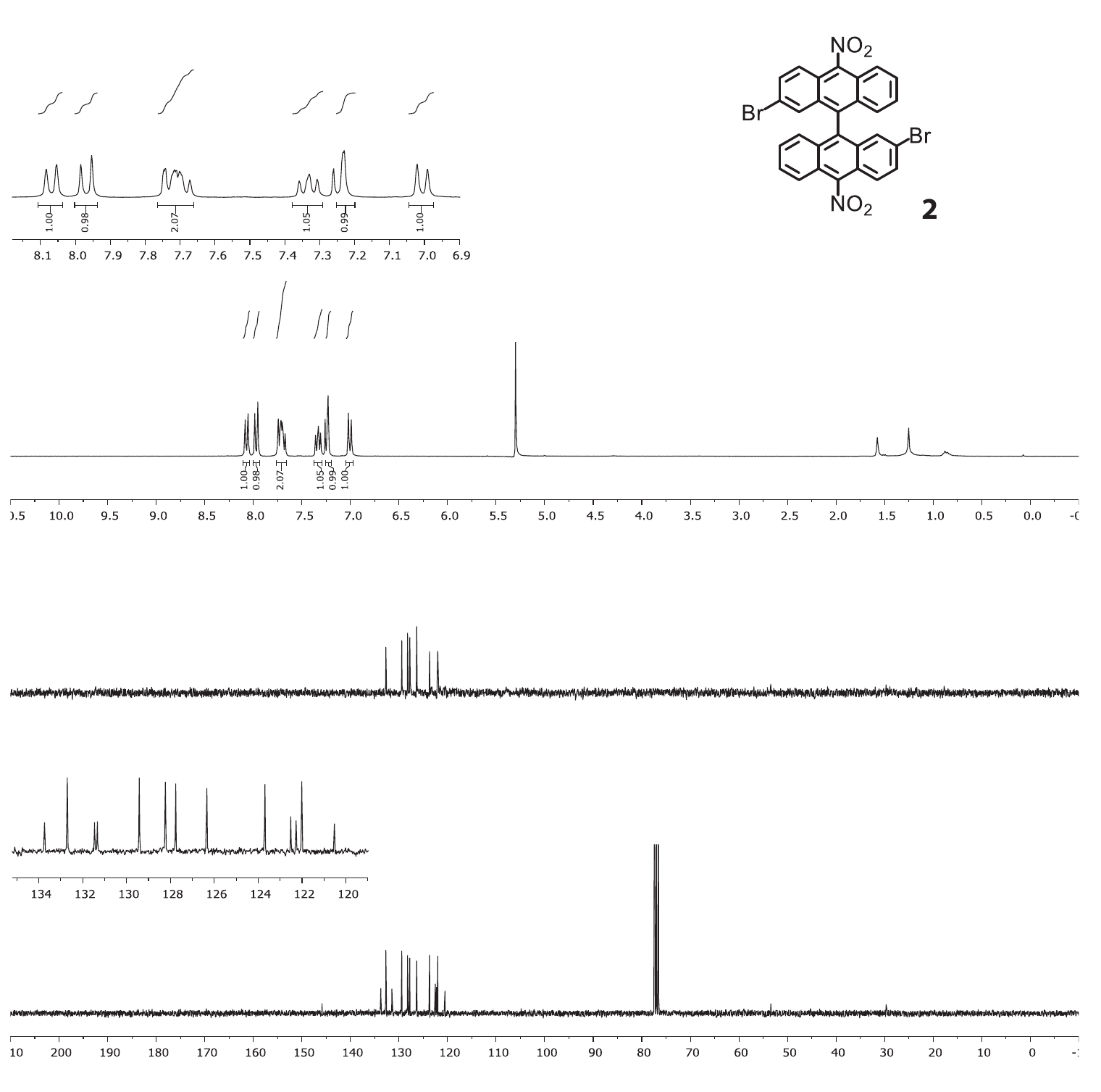}
\caption{$^{1}$H and $^{13}$C NMR spectra of compound \textbf{2}.}                                                                                 
\end{figure}

\begin{figure}[h]
\includegraphics{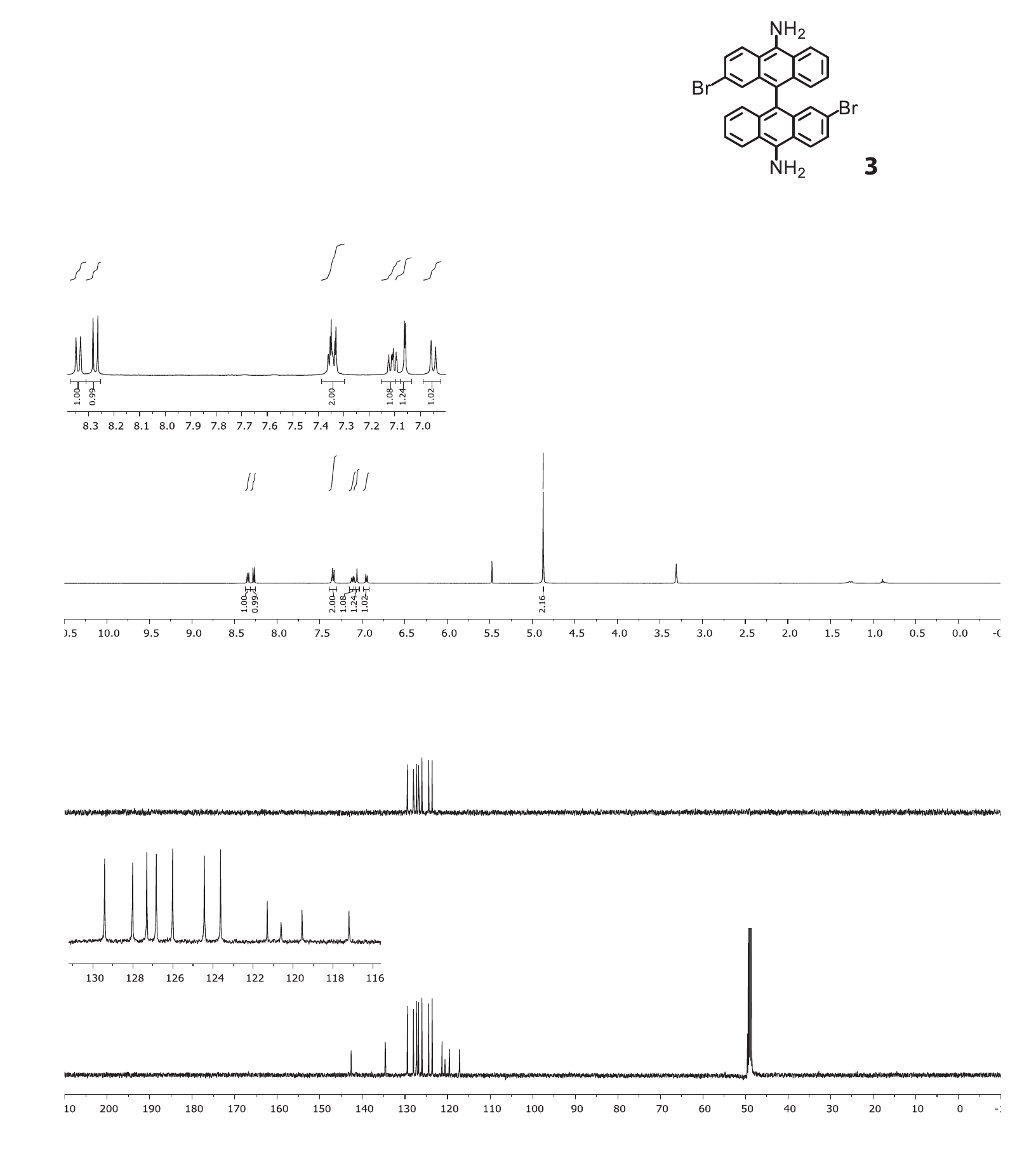}
\caption{$^{1}$H and $^{13}$C NMR spectra of compound \textbf{3}.}        
\end{figure}

\clearpage

 \subsection{Line profiles of NH$_{2}$-chGNRs and pristine chGNRs}
 Here we compare the line profiles accross the NH$_{2}$-chGNRs and pristine chGNRs. The line profile accross NH$_{2}$-chGNRs is around 0.5 nm wider than that of pristine chGNRs (Figure S5). 
\begin{figure}[htb]
\includegraphics{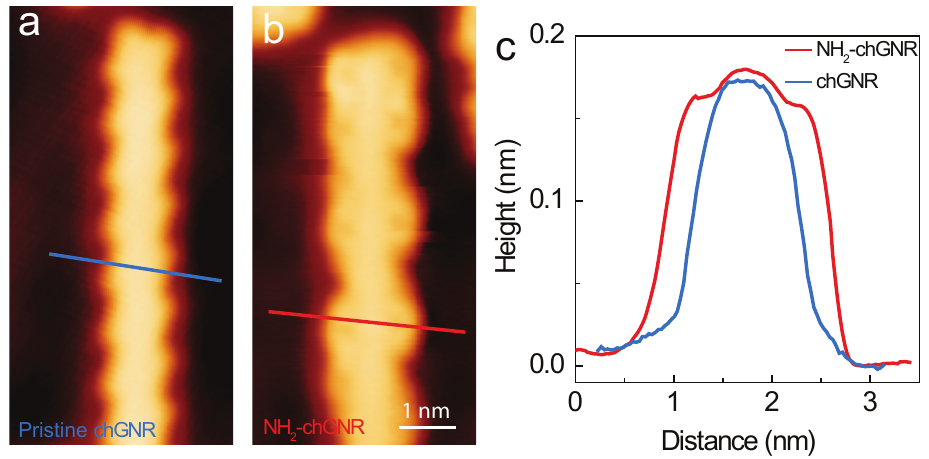}
\caption{Line profiles of NH$_{2}$-chGNRs and pristine chGNRs. \textcolor{black}{(a,b) Constant current STM images of pristine chGNRs and NH$_{2}$-chGNRs. The images share the same scale bar. (c), Line profiles across pristine chGNRs and NH$_{2}$-chGNRs with lines indicated in (a,b).} }                                                                                 
\end{figure}